\documentstyle[12pt]{article}

\begin{document}

\begin{titlepage}
\noindent
%
%
\hfill SLAC-PUB-95-6889  \\
\mbox{}
\hfill TTP95--21\footnote{The complete postscript file of this
preprint, including figures, is available via anonymous ftp at
ttpux2.physik.uni-karlsruhe.de (129.13.102.139) as
/ttp95-21/ttp95-21.ps or via www at
http://ttpux2.physik.uni-karlsruhe.de/cgi-bin/preprints/
Report-no: TTP95-21.}\\
\mbox{}
\hfill hep-ph/9505303\\

\vspace{.5cm}

\begin{center}
\begin{LARGE}
 {\bf
Tau Polarimetry with Multi-Meson States
 }
      \footnote{Work supported in part by the Department of Energy,
                                 Contract DE-ACO3-76SF00515
                                 and
                                 BMFT Contract 056KA93P6.}
\end{LARGE}

\vspace{.8cm}

\begin{large}
 J.H.~K\"uhn
\end{large}

\vspace{.5cm}
   Institut f\"ur Theoretische Teilchenphysik\\
   Universit\"at Karlsruhe, Kaiserstr. 12,    Postfach 6980,
   D-76128 Karlsruhe, Germany\\
\vspace{.5cm}
and \\
\vspace{.5cm}
   Stanford Linear Accelerator Center\\
   Stanford University, Stanford, CA 94309

\vspace{.5cm}

\begin{abstract}
\noindent
It is demonstrated that the analyzing power of
multi-meson final states in semileptonic $\tau$
decays with respect to the $\tau$ spin is equal and
maximal for all decay modes.
\end{abstract}

\vspace{2cm}

\centerline{(Submitted to Physical Review D)}

\vfill
\end{center}
\end{titlepage}

\renewcommand{\arraystretch}{2}

\noindent
The determination of the polarization of
$\tau$ leptons produced in $Z$
decays has lead to an important determination
of the $\tau$ coupling to
the $Z$ boson, rivaling those from the
forward-backward asymmetry
of leptons or from the left-right asymmetry
measured with longitudinally
polarized beams. It is well known that the
decay mode into a single
pion leads to optimal analyzing power which
expresses itself in an
angular distribution of pions from decays
of polarized $\tau$ 's of the form
\begin{equation}
dN\propto (1+ f \cos \theta)
\end{equation}
where $f=1$. The angle
between $\tau$ spin and pion momentum
is denoted by $\theta$.
Decays into the $\rho$ or $a_1$ meson or higher
excitations exhibit reduced analyzing power, e.g.
$f=(M^2  - 2 Q^2) / (M^2 + 2 Q^2)$
for a spin 1 final state of mass $Q$ \cite{Tsai}.
In \cite{KW,many}
it has been argued that significant analyzing
power can be recovered by
exploiting information encoded in the momenta
of the (pseudoscalar) mesons which are the actual decay
products and are observed in the experiment.
Detailed models have been used for the two-
and three-meson channels to
identify various angular distributions which
enhance the sensitivity.

In this brief comment we would like to
demonstrate explicitly that maximal
sensitivity can be recovered for {\it any}
multi-meson final state, once
the dynamics of the decay matrix element is known.
Ingredients are the knowledge of all meson momenta  and
information about the $\tau$ rest frame. The
latter is equivalent to reconstruction of the
actual direction of flight of the $\tau$ and can be
achieved in $e^+e^-$ experiments with the help of
vertex detectors \cite{impact}.

The argument is based on the observation
that the squared matrix element for semileptonic
$\tau$ can always be written (in the $\tau$
restframe) in the form
\begin{equation}
{\cal M}\propto 1 - \vec h \vec s .
\end{equation}
The $\tau$-spin direction is denoted by $\vec s$
and the polarimeter vector $\vec h$
is a function of all meson momenta. It has
length $\vec h=1$ and its direction therefore
gives the (negative) spin direction of the original $\tau$
with unit probability. This holds true for meson
final states only --- the spin information is
strongly diluted for leptonic $\tau$ decays as a
consequence of averaging over the electron spin
and neutrino momenta corresponding to a reduction
of the length of $\vec h$.
It is,  however, retained in the direction of the $\bar\nu_e$.
For top decays, the $e^+$ direction preserves the full
analyzing power \cite{top}.

The aim of this short comment is to demonstrate
that the norm of $\vec h$ is in fact equal to
one for all semileptonic decays.

The matrix element for the semileptonic decay
$\tau\to\nu_\tau + X$ can be written in the form
\begin{equation}
{\cal M} =\frac{G}{\sqrt2}\bar u(N) \gamma^\mu (1-\gamma_5)u(P)J_\mu,
\end{equation}
where $J_\mu\equiv \langle X| V_\mu - A_\mu| 0\rangle$ denotes the
matrix element of the $V-A$ current relevant for the specific final
state $X$. The vector $J_\mu$ depends in general on the momenta of all
hadrons. The squared matrix element for the decay of a $\tau$ with
spin $s$ and mass $M$ then reads
\begin{equation}
|{\cal M}|^2 = G^2 (\omega + H_\mu s^\mu),
\end{equation}
with
\begin{equation}
\omega = P_\mu (\Pi_\mu + \Pi^5_\mu)~~~~~~~~~~~~~~
H_\mu = \frac{1}{M}
(M^2 g_\mu^{\phantom{\mu}\nu} - P_\mu P^\nu) (\Pi_\nu + \Pi^5_\nu)
\end{equation}
and
\begin{equation}
\Pi_\mu = 2\left [ (J^*\cdot N) J_\mu + (J\cdot N) J^*_\mu -
(J^*\cdot J) N_\mu\right], ~~~~~~~
\Pi^5_\mu =2\, {\rm Im}\,
\epsilon_\mu^{\phantom{\mu}\nu\rho\sigma} J_\nu^* J_\rho N_\sigma .
\end{equation}

This formula was derived in \cite {KW} and
constitutes the basis for the simulation of spin
effects in TAUOLA \cite{tauola}.

In the
$\tau$-rest frame, the function $\omega$ coincides
with the time component of the four vector
$\Pi_\mu + \Pi^5_\mu$ (multiplied with $M$) and the vector $\vec
H$ with its space component (multiplied with $M$). The
assertion that $|\vec h| = |\vec H/\omega| =1$ is
therefore equivalent to the statement that $\Pi_\mu +\Pi^5_\mu$
is null-vector. A simple calculation demonstrates that
\begin{equation}
\Pi^5_\mu\Pi^\mu =0,~~~~~~~~~ \Pi^5_\mu \Pi^{5\mu}
= -\Pi_\mu \Pi^\mu
\end{equation}
and hence
\begin{equation}
(\Pi_\mu + \Pi_\mu^5) (\Pi^\mu + \Pi^{5\mu}) = 0
\end{equation}
which proves our assertion and gives, at the same time,
a prescription of how to construct the direction of the
spin on an event-by-event basis.
Obviously, in order to perform this analysis, all
hadron momenta must be measured and the dependence of the
current $J$ on these momenta known --- either from
theoretical considerations or from fits to
experimentally measured distributions.

\vspace{5ex}
\noindent
{\bf Acknowledgments}

\noindent
The author would like to thank
D. Atwood for discussions,
 the SLAC theory group for hospitality, and the
Volkswagen-Stiftung grant I/70 452 for generous support.


\begin{thebibliography}{99}

\bibitem{Tsai} Y.S. Tsai, Phys.\ Rev.\ D 4 (1971) 2821.

\bibitem{KW} J.H. K\"uhn and F. Wagner,
             Nucl.\ Phys.\ B 236 (1984) 16.
\bibitem {many} K. Hagiwara, A.D. Martin and D. Zeppenfeld,
                 Phys.\ Lett. B 235 (1989) 198;\\
                A. Rou\'ge, Z.\ Phys. C 48 (1990) 77;\\
        J.H. K\"uhn and E. Mirkes, Phys.\ Lett.\ B 286 (1992).

\bibitem {impact}J.H. K\"uhn, Phys.\ Lett.\ B 313 (1993) 458.

\bibitem{top} J.H. K\"uhn and K.H. Streng, Nucl.\ Phys.\ B 198
               (1982) 71;\\
              M. Je\.zabek and J.H. K\"uhn, Nucl.\ Phys.\ B 320
                (1989) 20.

\bibitem {tauola}S. Jadach, J.H. K\"uhn and Z. Wa\c s,
    Comput.\ Phys.\ Commun.\ 64 (1991) 275;\\
    M. Je\.zabek, Z. Wa\c s, S. Jadach and J.H. K\"uhn,
    {\em ibid.}, 70 (1992) 69;\\
 S. Jadach, Z. Wa\c s, R. Decker and J.H. K\"uhn,
 {\em ibid.}, 76 (1993) 361.


\end{thebibliography}
\end{document}